\documentclass[twocolumn]{aastex61}
\usepackage{graphicx}
\usepackage{color}

\begin{document}

\title{High Resolution Spectroscopy using Fabry Perot Intereferometer Arrays: An Application to Searches for O$_{2}$ in Exoplanetary Atmospheres}
\author{Sagi Ben-Ami}
\affiliation{Harvard-Smithsonian Center for Astrophysics, 60 Garden Street, Cambridge, MA 02138, USA}
\author{Mercedes L${\rm  \acute{o}}$pez-Morales}
\affiliation{Harvard-Smithsonian Center for Astrophysics, 60 Garden Street, Cambridge, MA 02138, USA}
\author{Juliana Garcia-Mejia}
\affiliation{Harvard-Smithsonian Center for Astrophysics, 60 Garden Street, Cambridge, MA 02138, USA}
\author{Gonzalo Gonzalez Abad}
\affiliation{Harvard-Smithsonian Center for Astrophysics, 60 Garden Street, Cambridge, MA 02138, USA}
\author{Andrew Szentgyorgyi}
\affiliation{Harvard-Smithsonian Center for Astrophysics, 60 Garden Street, Cambridge, MA 02138, USA}
\correspondingauthor{Sagi Ben-Ami}
\email{sagi.ben-ami@cfa.harvard.edu}

\begin{abstract}
We present a novel implementation for extremely high resolution spectroscopy using custom-designed Fabry Perot Interferometer (FPI) arrays. For a given telescope aperture at the seeing limited case, these arrays can achieve resolutions well in excess of ${\rm R\sim10^5}$ using optical elements orders of magnitude smaller in size than standard echelle spectrographs of similar resolution. We apply this method specifically to the search for molecular oxygen in exoplanetary atmospheres using the ${\rm O_2}$ A-band at 0.76 ${\rm \mu m}$, and show how a FPI array composed of $\sim10$ etalons with parameters optimized for this science case can record ${\rm R=3-5\,\cdot10^5}$ spectra covering the full ${\rm O_2}$ A-band. Using simulated observations of the atmosphere of a transiting nearby Earth-like planet, we show how observations with a FPI array coupled to a long-slit spectrograph can reduce the number of transit observations needed to produce a ${\rm 3\sigma}$ detection of ${\rm O_2}$ by $\sim30\%$ compared to observations with a ${\rm R=10^5}$ echelle spectrograph. This, in turn, leads to a decrease in an observing program duration of several years. The number of transits needed for a ${\rm 3\sigma}$ detection can be further reduced by increasing the efficiency of FPI arrays using dualons (an etalon with a buried reflective layer), and by coupling the FPI array to a dedicated spectrograph optimized for the ${\rm O_2}$ A-band.
\end{abstract}

\section{Introduction}
High Resolution (HiRes) spectroscopy is a powerful tool for the detection and characterization of exoplanets. HiRes spectrographs are routinely used to measure exoplanet masses via  precision radial velocity techniques \citep[{\rm PRV}; \textit{e.g.,}][]{Fischer2016}. With new instruments and analysis techniques regularly breaking stability and precision limits, HiRes spectrographs now posses the capability to detect terrestrial exoplanets in the habitable zone of nearby, low mass host stars \citep[\textit{e.g.,}][]{Anglada2016}. Using transmission and emission spectroscopy methods, HiRes spectrographs have also successfully detected molecules such as ${\rm CO}$, ${\rm CH_4}$, and ${\rm H_2O}$ in the atmospheres of gas giant exoplanets \citep[\textit{e.g.,}][]{Snellen2010,Brogi2012,Rodler2012,Birkby2017,Evans2017}. 

Upcoming missions like the NASA Transiting Exoplanet Survey Satellite, \citep[TESS;][]{Ricker2016} and ESA CHaracterising ExOPlanet Satellite \citep[CHEOPS;][]{Cessa2017} will discover and photometrically characterize dozens of Earth-like planets in the next decade \citep[\textit{e.g.,}][]{Sullivan2015}. These exoplanets will be bright and amenable to detailed PRV and transmission spectroscopy observations with the next generation of Extremely Large Telescopes (ELTs). Studying these planets atmospheres in search for potential biomarkers like ${\rm O_2}$ is the next natural step \citep{Snellen2013}.

We have recently performed a series of simulations aimed at establishing the optimal spectral resolution and the optimal combination of wavelength bands to detect ${\rm O_2}$ in the transmission spectrum of an earth-like planet (L${\rm  \acute{o}}$pez-Morales et al. 2018; In Prep). 
Our study indicates that observing at resolutions of $R\sim3-5\cdot10^5$ will increases the signal-to-noise of the ${\rm O_2}$ signal, and reduce the observing time necessary for a detection. Such a high resolution is above the typical resolution of seeing limited HiRes spectrographs \citep[${\rm R\sim10^5}$; \textit{e.g.,} HARPS, EXPRES;][]{Cosentino2012,Jurgenson2016}.  
Telluric lines in the earth atmosphere are resolved at resolutions well below ${\rm R\sim10^5}$. These lines have a Voigt profile, with a broad component associated with pressure broadening at the lower layers of the earth atmosphere \citep[\textit{e.g.,}][]{Chance2017}. Due to refraction effects, transmission spectroscopy does not sample the lower layers of an exoplanet atmosphere, and higher resolving power is required to fully sample the molecular oxygen absorption lines.

Most astronomical HiRes spectrographs are echelle spectrographs. These achieve high spectral resolutions by dispersing light with a low line density ruled grating blazed for high diffraction orders. The spectral resolution of an echelle spectrograph in the seeing limited case is given by \citep{Schroeder2000}:
\begin{equation}
{\rm R} = {\rm \frac{2d_{col} tan\delta}{\psi d_{tel}} }
\end{equation}
\noindent where  ${\rm d_{col}}$ is the diameter of the collimated beam dispersed by the echelle grating (effectively the diameter of the spectrograph collimator), ${\rm d_{tel}}$ is the diameter of the telescope primary mirror, ${\rm \psi}$ is the diameter of the seeing disk in radians (\textit{i.e.,} effective PSF), and ${\rm \delta}$ is the grating blaze angle (with $\tan\delta$ technologically limited to values $\leq4$). The scaling of resolution with inverse PSF and telescope diameter is a consequence of Lagrange invariance \citep[\textit{e.g.,}][]{Schroeder2000}. Eq. 1 indicates that at seeing limited conditions, achieving high resolutions requires a large format instrument - increasing its complexity and cost. For extremely large telescopes, like the ones needed to detect ${\rm{O}_2}$ absorption in exoplanets atmospheres \citep{Rodler2014}, such instruments become unfeasible. 
   
Two methods that have been proposed to achieve HiRes with large telescopes while maintaining a reasonable instrument size are pupil slicing \citep[\textit{e.g.,}][]{Conconi2013,Seifahrt2016} and adaptive optics \citep[AO; \textit{e.g.,}][]{Crepp2016}. 
In the pupil slicing method, one reduces the effective size of the PSF by slicing the pupil of the telescope to sub-apertures. As an example, in high resolution modes, The GMT Consortium Large Earth Finder (G-CLEF) slices the Giant Magellan Telescope (GMT) aperture into seven circular sub-apertures. Inputs from the sub-apertures are arranged along the spectrograph slit. This is equivalent to feeding a single spectrograph with seven telescopes simultaneously, each one with 1/7 of the GMT aperture.
The 6-fold symmetry of the GMT pupil allows for minimum throughput loss \citep{Ben-Ami2016}. While the pupil slicing technique has been implemented in various instruments \citep[\textit{e.g.,}][]{Conconi2013,Seifahrt2016}, it suffers from two main drawbacks when increasing the number of sub-apertures: (\textit{i.}) Slit losses in pupil slicers are often large; (\textit{ii.}) The slit length must be increased to accommodate all sub-apertures, further complicating the instrument design and potentially limiting the number of diffraction orders that can be imaged simultaneously on the instrument detector.

AO may be used to drastically reduce seeing disk size and thus increase resolution without increasing instrument size. If AO could reduce the telescope point spread function to the $10\mu m$ scale, single mode optical fibers (SMFs) might be use to feed spectrographs. This will have the added benefit of vastly improving wavelength scale stability \citep{Crepp2016}. Recently, \cite{Bechter2016} and \cite{Jovanovic2016} have demonstrated AO-SMF coupling efficiencies of up to $25\%$ at the NIR. While this is a significant step in enabling extreme high-resolution spectrographs for large aperture telescopes, the demonstrated efficiencies are still somewhat lacking for a photon starving measurement such as ${\rm O_2}$ detection in transmission spectroscopy. In addition, reducing the PSF to a $10\mu$m scale with an AO system has not been demonstrated at the ${\rm O_2}$ A-band around $760\,$nm, the preferred detection band
\citep[][L{$\rm\acute{o}$}pez-Morales et al. 2018, In Prep]{Rodler2014}.

An innovative approach for increasing the resolving power of a spectrograph is Externally Dispersed Interferometry \citep[EDI; ][]{Erskine2003,Muirhead2011,Erskine2016}. In this method, an external Michelson interferometer images fringes to a spectrograph slit. The fringes, with the input spectrum, heterodyne fine spectral features into a low spatial frequency moire pattern. The heterodyning is then inverted numerically to recover detailed spectral information at extremely high resolution. EDI has the capability of significantly increasing resolving power over that obtainable with echelle spectrograph by extremely economical means. However, full spectroscopic information recovery requires repeated observations  at multiple phase delays. EDI instrumentation lowers the cost of the instrument at a given resolution, yet this economy is made at the expense of increased telescope time. These savings becomes less and less attractive as the value of telescope time grows with increasing aperture. It is also to be noted that exoplanets transits themselves are quite time constrained, and so a requirement for repeated observations is sub-optimal for transmission spectroscopy during transits.

Here we propose a novel instrumental approach to increase the resolving power of spectrographs on ELTs. The scientific motivation guiding us is the detection of ${\rm O_2}$ in the atmosphere of Earth-like planets. Our concept instrument is based on the transmissive and reflective properties of Fabry Perot Interferometers (FPIs) chained together. Our concept uses an external spectrograph working at a significantly lower resolution to separate the interference orders intrinsic to Fabry Perots. In Section 2 we describe our instrumental method and review the transmission and reflection properties of a single FPI. We then apply those results to model the response of an n-fold array. The section ends with a discussion of interference order dispersion and the role of an external spectrograph. In Section 3 we discuss practical considerations involved in the implementation and performance of an instrument based on the methodology described in Section 2. Section 4 presents simulation results for observations done with a FPI array instrument designed to detect {$\rm O_2$} in the atmosphere of an earth-like planet via transmission spectroscopy. Discussion, summary, and future direction of this work are given in Section 5.

\section{Instrumental Method}
A typical FPI consists of two flat, parallel, partially reflective optical surfaces facing each other and separated by a distance ranging from microns to centimeters. The emergent beam on both sides of the FPI constitute a high multiplicity beam interferometer, which  makes it possible to achieve extremely high resolving power in extremely compact formats. For the purpose of these discussions, we assume the geometry of the FPI is planar, although FPIs with curved surfaces offer advantages for some applications. One commonly used class of FPIs is etalons, in which the medium between the two flat surfaces is typically high homogeneity optical glass or crystal.

\begin{figure}[]
\centering
{\includegraphics[width=1\columnwidth]{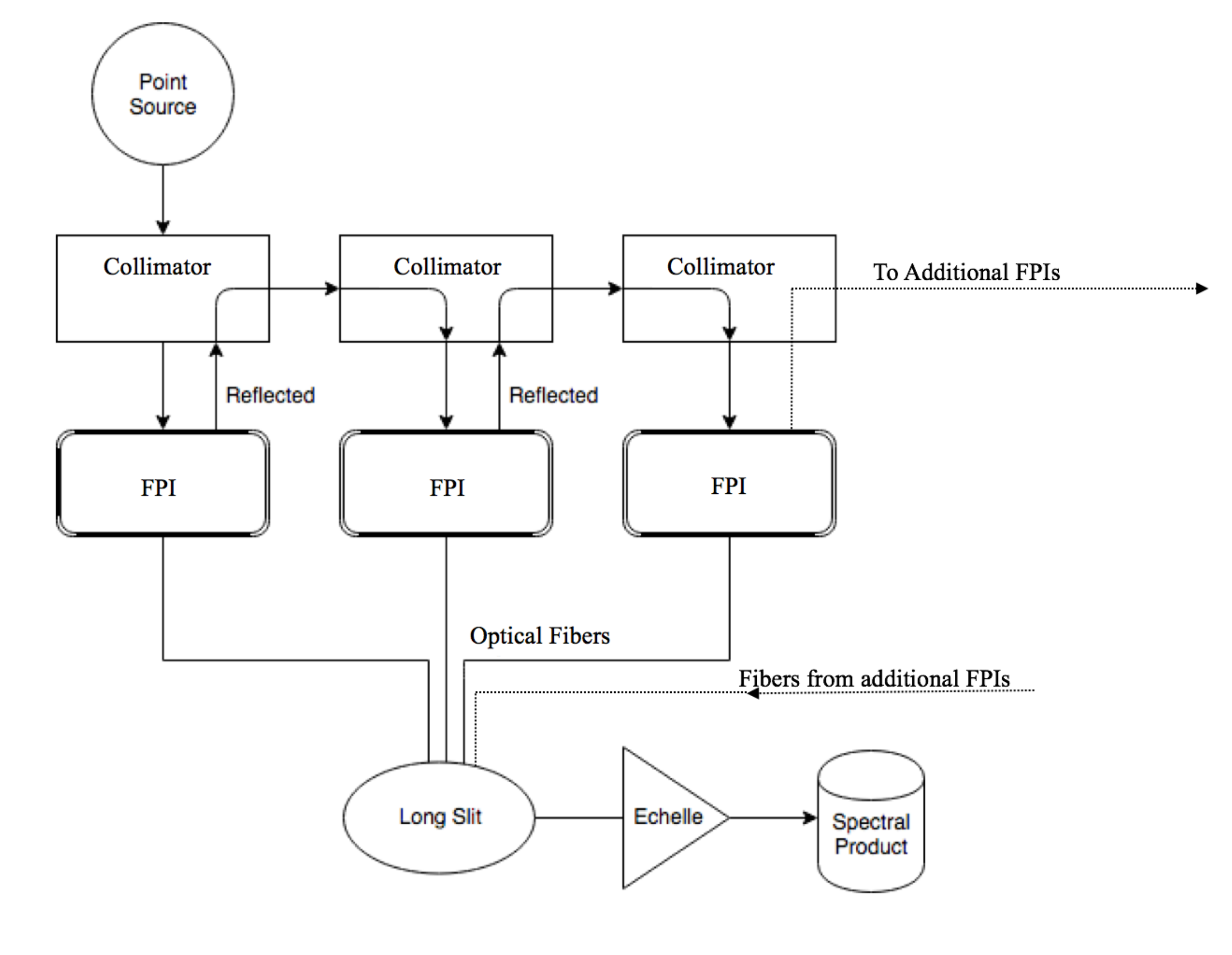}} 
\caption{Block Diagram of our FPI-based method for extreme HiRes spectroscopy. A series of FPIs generate transmission combs with the desired resolving power. The sampling frequency is increased by a factor equals to the number of FPIs in the chain. An external long-slit spectrograph separates the overlapping orders from each FPI in the chain. For clarity, we show a short chain of three FPIs.}
\label{fig:BlockLayout}
\end{figure}

In the following subsections we describe in detail our method of using etalons to produce spectra with resolving power and sampling frequency well in excess of ${\rm R\sim10^5}$. Our description focuses on recording HiRes spectra centered on the ${\rm O_2}$ A-band, but the same technique is applicable to other wavelengths.
Figure \ref{fig:BlockLayout} shows a block diagram of our method from the telescope image plane to the final spectrum.  

\subsection{Properties of a single etalon}
The transmission and reflection intensities from a single etalon, neglecting absorption at the reflective surfaces\footnote{These will have negligible effects of less than $1\%$ to the system throughput and spectral resolving power FWHM - assuming high quality coatings common in modern precision optics (based on a discussion with leading vendors).},  are given by \cite{Vaughan1989}:
\begin{equation}
I_T=\frac{T^2}{(1-R)^2(1+F\sin^2(\frac{\phi}{2}))}=\frac{T^2}{(1-R)^2A(\phi)}   
\end{equation}
\begin{equation}
I_R=\frac{F\sin^2(\frac{\phi}{2})}{1+F\sin^2(\frac{\phi}{2})}=\frac{F\sin^2(\frac{\phi}{2})}{A(\phi)}
\end{equation}
\noindent where $\phi$ is the phase lag in a double pass through the etalon, R is the Surface Reflectivity Intensity (SRI), and T is the Surface Transmissivity Intensity (STI). $A(\phi)$ is the Airy function and $F\equiv\frac{4R}{(1-R)^2}$ is the coefficient of finesse.
\begin{figure}[]
\centering
{\includegraphics[width=1\columnwidth]{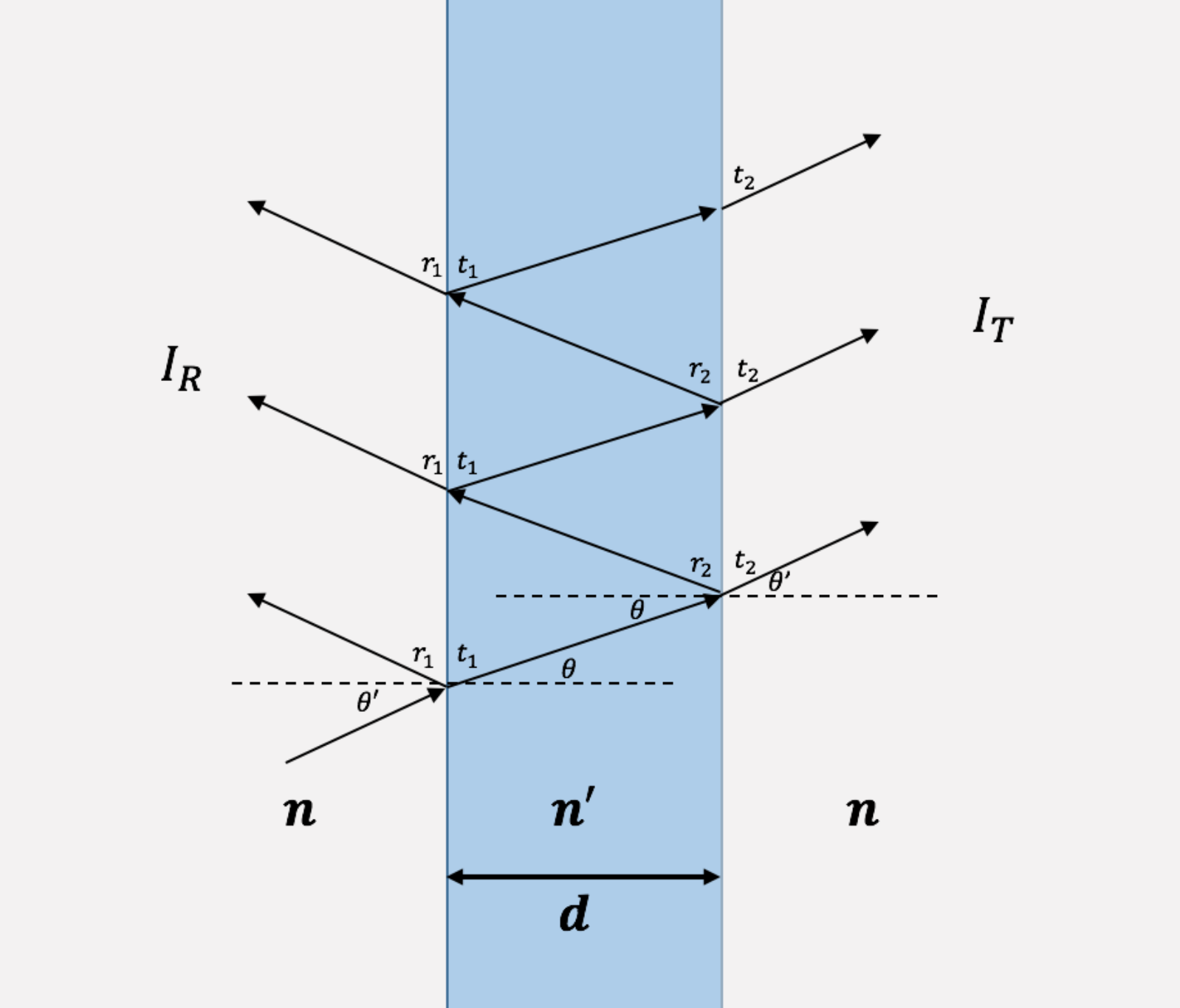}} 
\caption{Rays Transmission/Reflection through a Fabry Perot Interferometer.}
\label{fig:FPI_Layout}
\end{figure}

\begin{figure}[]
\centering
{\includegraphics[width=1\columnwidth]{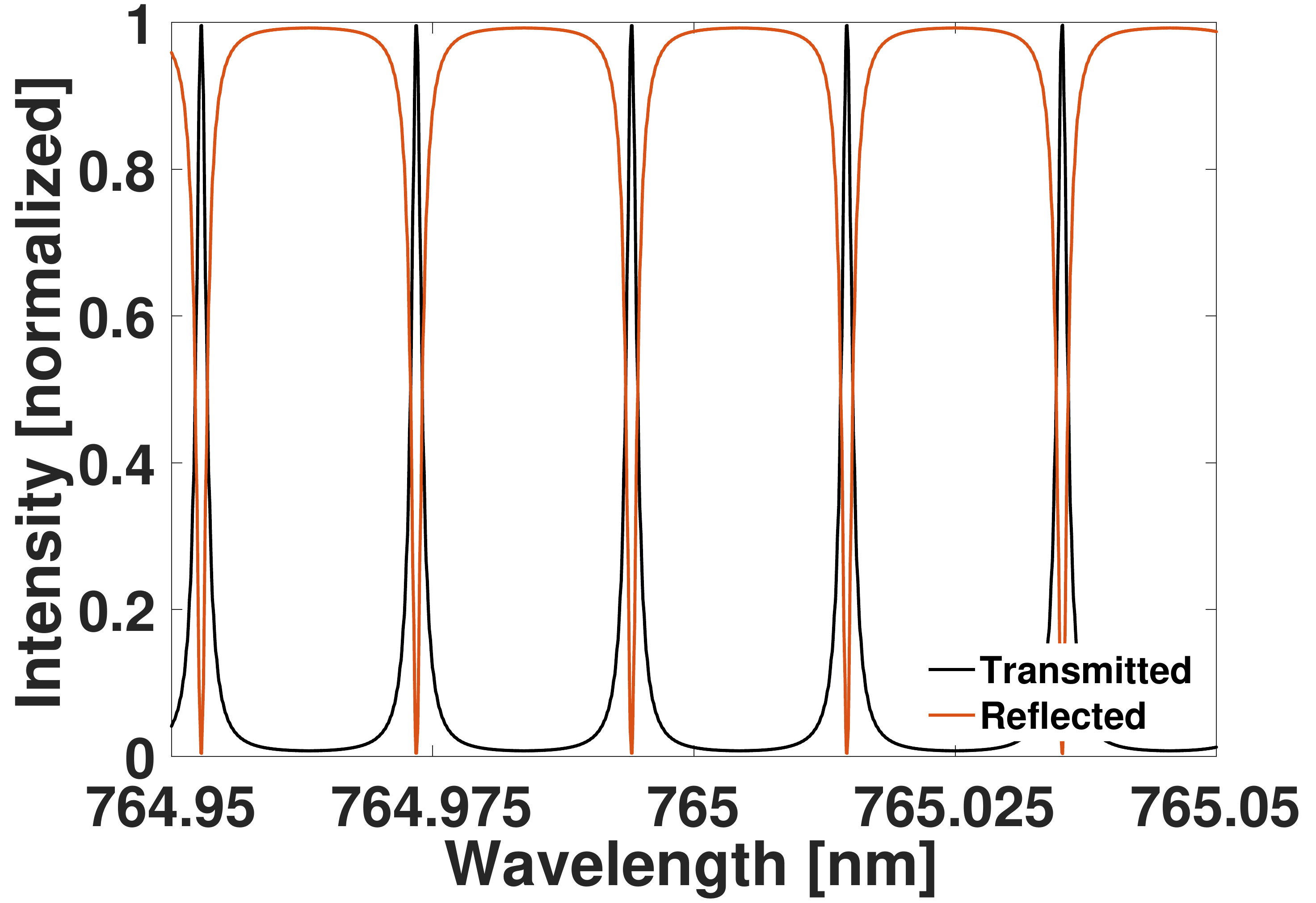}} 
\caption{Theoretical transmission/reflection Intensity of a single etalon, with mechanical and optical parameters as specified in Table 1. The contrast in the transmitted/reflected beam is $\frac{I_{max}}{I_{min}}>200$ (\textit{i.e.,} $23$db).}
\label{fig:singleFPIResponse}
\end{figure}

The phase lag depends on the separation between the reflective surfaces $\rm d$, the index of refraction of the medium ${\rm n'}$, and the the angle of incidence of the collimated beam with respect to the $2^{\mathrm{nd}}$ reflective surface $\rm{\theta}$, see Fig. \ref{fig:FPI_Layout}:
\begin{equation}
{\rm \phi} = {\rm \frac{2\pi}{\lambda} {\rm ~2d} {\rm n'} {\rm \cos\theta}}
\end{equation}
Figure \ref{fig:singleFPIResponse} shows the reflection and transmission curves of a single etalon with the parameters given in Table 1. Maximum transmitted intensity peaks occur at wavelengths satisfying the condition 
\begin{equation}
{\rm m\lambda}={\rm 2dn' \cos\theta}  
\end{equation}
\noindent
where $m$, an integer, is the interference order. The transmission efficiency drops rapidly between peaks, resulting in a comb-like transmission curve. For the case presented, the contrast is above $\frac{I_{max}}{I_{min}}>200$ (\textit{i.e.,} $23$db).

Adopting a resolving power definition similar to the Rayleigh Criterion, the spectral resolving power of an etalon is given by \citep{Vaughan1989}:
\begin{equation}
{\rm R_0}=\frac{\rm \lambda}{\rm \delta\lambda}=\frac{\rm 2\pi m}{\rm \delta\phi}=\frac{\pi m\sqrt{F}}{2}
\end{equation}
For the parameters listed in Table 1, we obtain a spectral resolution of ${\rm R_0 = 9.2\cdot10^5}$ for $\lambda=765\,nm$ at $\rm{m}=3.75\cdot10^4$.
Such a resolution will allow us to fully resolve the ${\rm O_2}$ A-band of an earth-like planet atmosphere (L${\rm  \acute{o}}$pez-Morales et al. 2018; In Prep). The analysis does not include imperfections in the plates or facets, and imperfect collimation effects, the impacts of are discussed in Section 3. 

The sampling frequency of a single etalon is dictated by its FSR:
\begin{equation}
  {\rm \Delta\lambda} = \frac{\rm \lambda^2}{\rm 2dn'\cos\theta}\longrightarrow\frac{\rm \lambda}{\Delta\lambda}=\frac{\rm 2dn'\cos\theta} {\rm \lambda} = {\rm m}
\end{equation}
From the FSR and spectral resolving power equations we see that the order of interference at a specific peak wavelength dictates the FSR at that wavelength, while spectral resolution is governed by both the interference order and the FPI coefficient of finesse ($F$). 

A figure of merit for a FPI quality is its finesse, defined as
\begin{equation}
  \mathcal{F}=\frac{\pi}{2arcsin(\frac{1}{\sqrt{F}})}
\end{equation}
While FPIs with a $\mathcal{F}>100$ are routinely produced, yielding resolving powers well in excess of ${\rm  R\sim10^6}$, the example etalon is of modest quality, with a finesse of $\mathcal{F}=24.56$.

The discussion above illustrates how a single etalon allows us to achieve very high spectral resolving power, but at a significantly lower sampling frequency, and so it cannot record a full continuous spectrum. Furthermore, the beam transmitted by an etalon will have overlapping interference orders, and additional dispersion is required for unambiguous spectral discrimination. The first limitation can be circumvented by using an array of etalons, as described in Section 2.2. The second limitation can be overcome by using a long-slit spectrograph to separate interference orders, as described in Section 2.3.

   {\centering
   \begin{deluxetable}{ccc}
   \tablecaption{Etalon Parameters}
   \tablehead{\colhead {Parameter} & \colhead {Name} & \colhead {Value}}
       \startdata     
       ${\rm d}$ & Spacing / Thickness & $9770\,\mu m$ \\
       ${\rm n'}$ & Index of Refraction & 1.4539 ( @ $\lambda=765\,$nm) \\
       ${\rm \theta}$ & Angle of Incidence &  $0.07^{\circ}$ \\
       $R_1, R_2$ & Surface Refl. Intensity & $R_1=R_2=0.88$ \\
       $T_1, T_2$ &  Surface Trans. Intensity & $T_1=T_2$ \\
       $\mathcal{F}$ & Finesse & $24.56$  \\
       $F$ &  Coefficient of Finesse & $244.44$ \\ 
       \enddata
       \label{tab:etalonParameters}
   \end{deluxetable}}

\subsection{The response of a FPI array}
\begin{figure}[]
\centering
{\includegraphics[width=1\columnwidth]{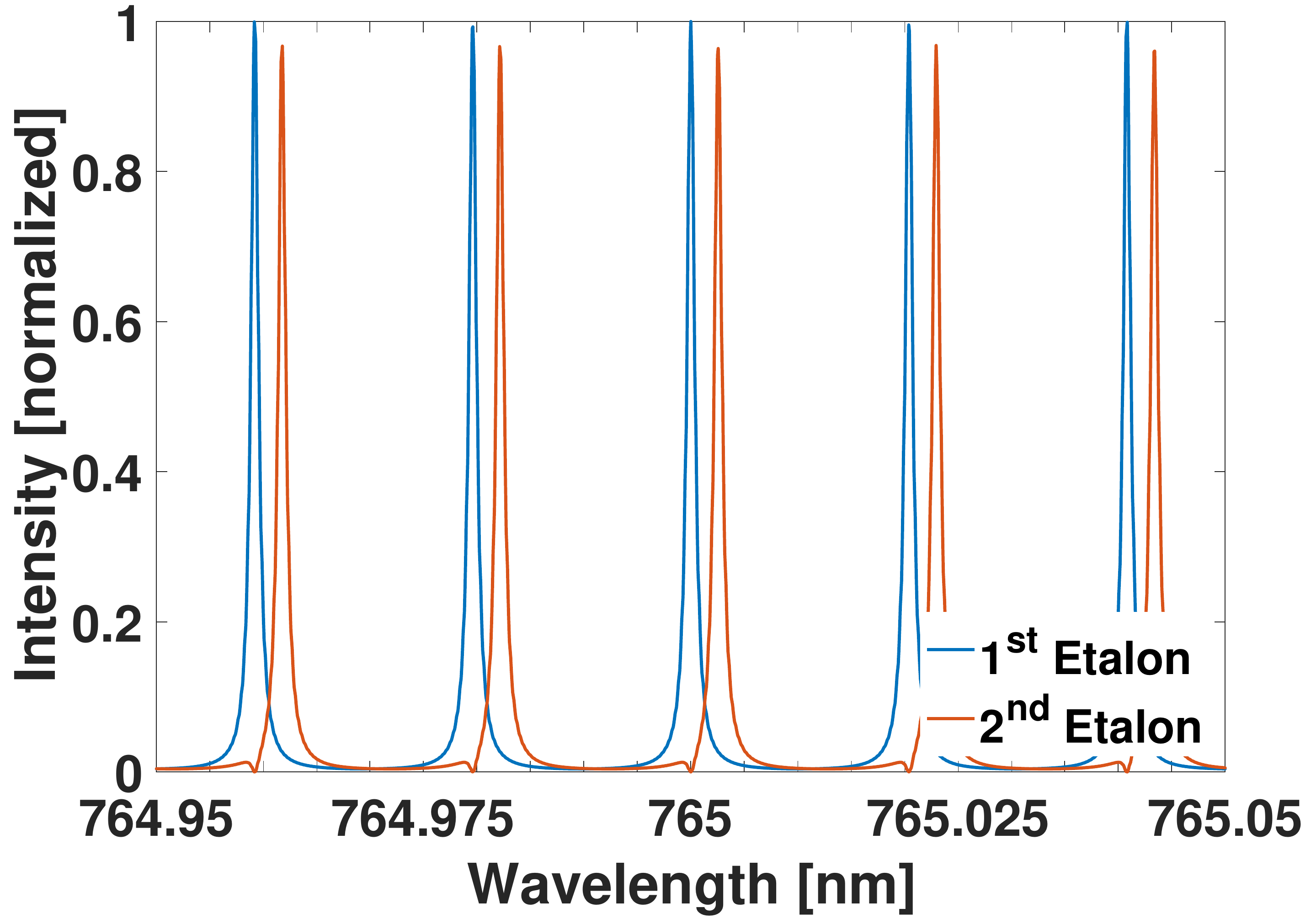}} \\
{\includegraphics[width=1\columnwidth]{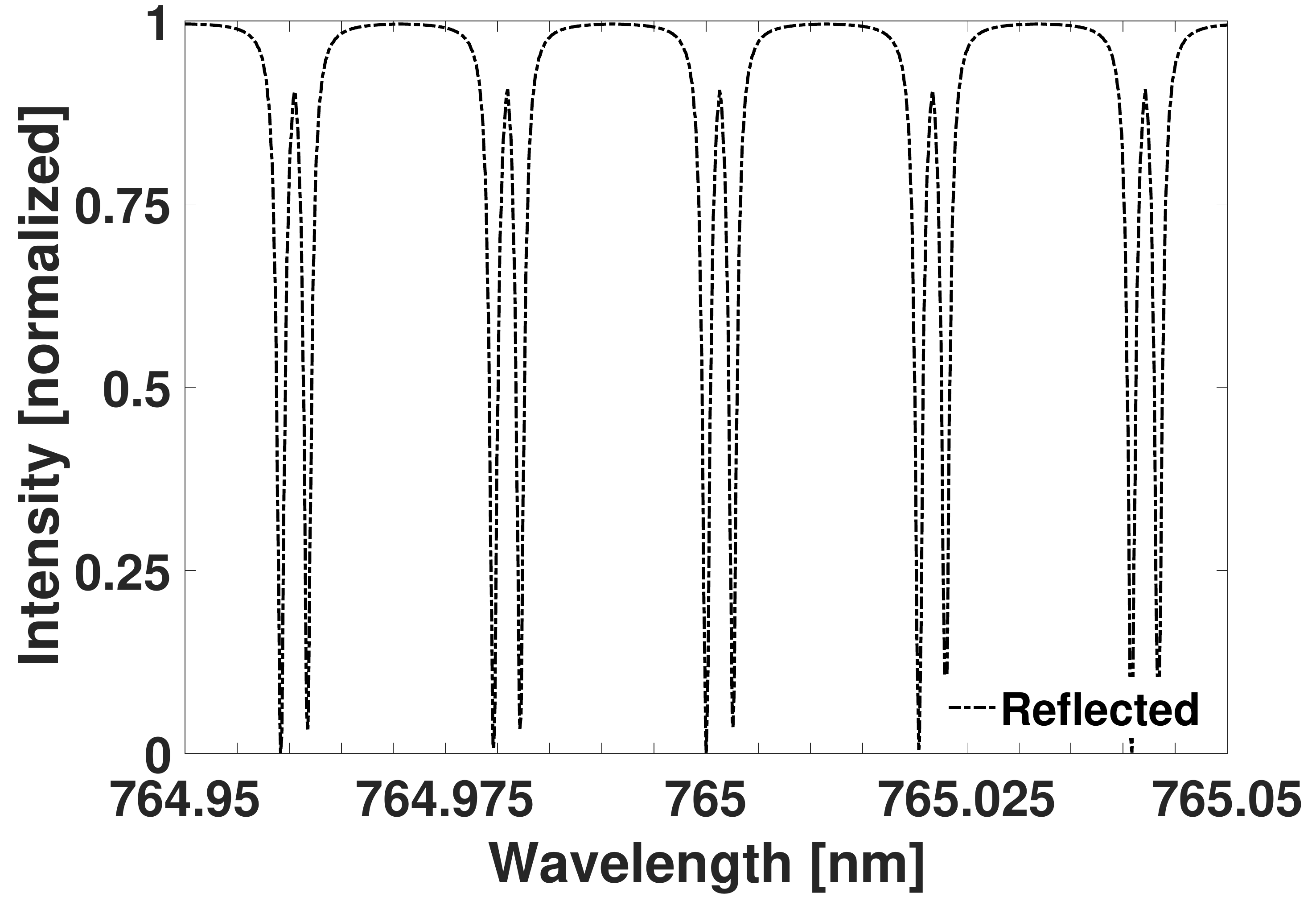}}  
\caption{Transmission (top) and reflection (bottom) response of two chained etalons, with the beam reflected from the first etalon directed to the second etalon. The two etalons differ in their thickness by $33\,$nm, so that for a specific order $\lambda_{FPI_2}^{peak}=\lambda_{FPI_1}^{peak}+{\lambda_{FPI_1}^{peak}}/{\rm R_{inst}}$, where ${\rm R_{inst}}=3\cdot10^5$ is the target spectral resolving power of the instrument.}
\label{fig:twoFPIResponse}
\end{figure}

\begin{figure}[]
\centering
{\includegraphics[width=1\columnwidth]{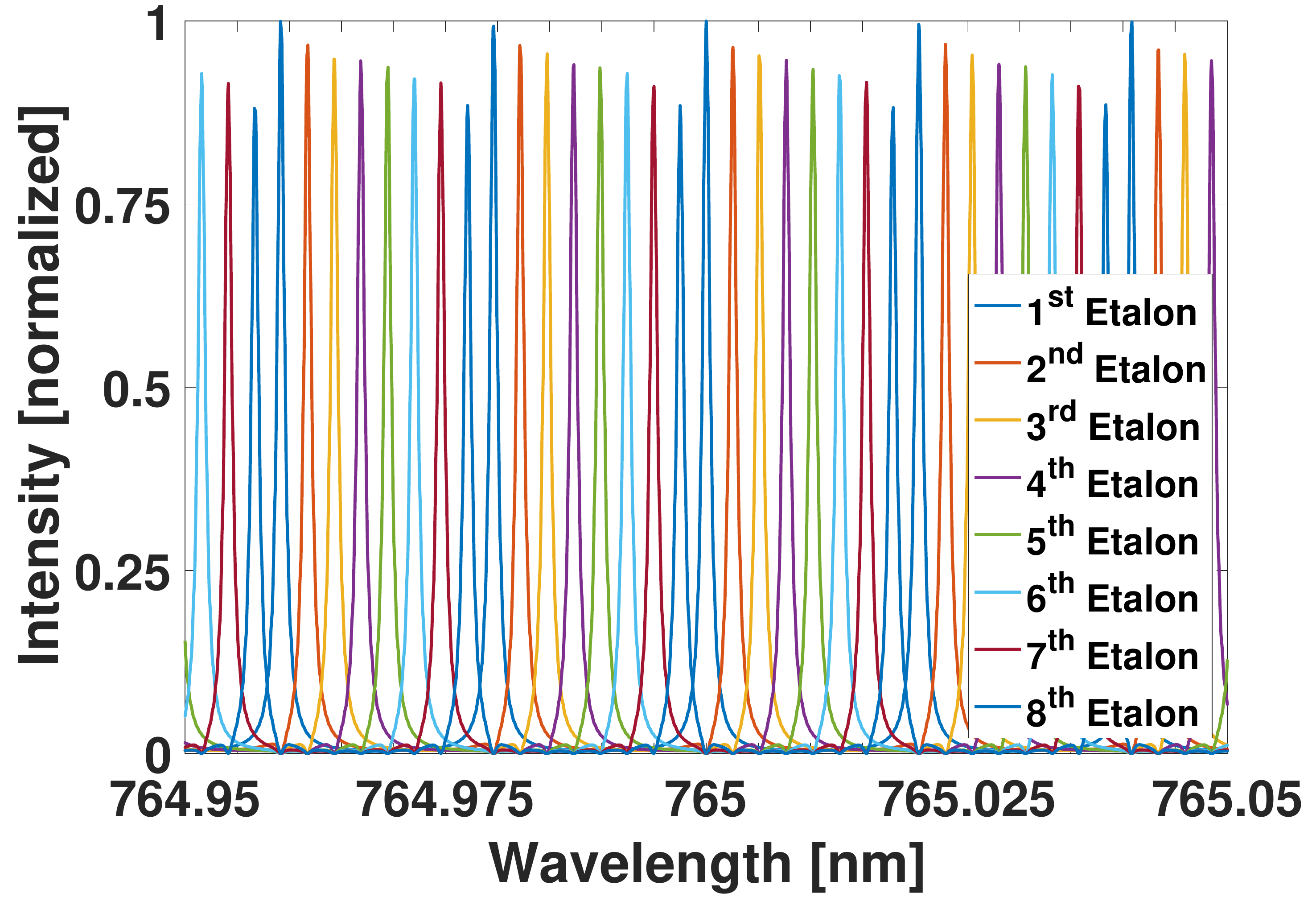}} 
\caption{Transmission comb of an 8-fold FPI array. Each etalon has a resolving power of $\frac{\lambda}{\delta\lambda}\sim9.2\cdot10^5$, and sampling frequency of $\frac{\lambda}{\Delta\lambda}=3.75\cdot10^4$. The overall result is a resolution of ${\rm R_{inst}}\sim3\cdot10^5$.}
\label{fig:FPIArrayResponse}
\end{figure}

A HiRes spectra with spectral resolving power and sampling frequency of similar magnitude can be recorded using a FPI array of multiple etalons, where the beam reflected from one etalon is redirected into the next etalon in the array, see Figure \ref{fig:BlockLayout}. 
We set the etalons thicknesses such that for a similar interference order 
\begin{equation}
    {\rm \lambda_{FPI_i}^{peak}}={\rm \lambda_{FPI_{i-1}}^{peak}}+{\rm \frac{\lambda_{FPI_{i-1}}^{peak}}{\rm R_{inst}}}
\end{equation}
where ${\rm \lambda_{FPI_i}^{peak}}$ and ${\rm \lambda_{FPI_{i-1}}^{peak}}$ are the central wavelengths of transmission peaks for two consecutive etalons, ${\rm i}$ and ${\rm i-1}$ in the array, and ${\rm R_{inst}}$ is the target resolution of the instrument, which we require to be equal to the instrument sampling frequency. All other etalon parameters are identical between etalons in the array. Figure~\ref{fig:twoFPIResponse} shows the transmission and reflection intensity profiles for two  adjacent etalons in an array assuming the parameters in Table 1. The thickness of the two etalons differs by ${\rm \Delta d=33\,nm}$, given a target resolution of ${\rm R_{inst}\sim3\cdot10^5}$. For the example FPI array targeting detection of ${\rm O_2}$ in the A-band, we can achieve a resolution of ${\rm R\sim3\cdot10^5}$ with 8 etalons, each with a theoretical spectral resolution of ${\rm\frac{\lambda}{\delta\lambda}\sim9.2\cdot10^5}$, and a sampling frequency of ${\rm \frac{\lambda}{\Delta\lambda}=3.75\cdot10^4}$. The resulting transmission comb of such an array  is illustrated in Figure \ref{fig:FPIArrayResponse}. 

While thickness variation of $30\,nm$ between adjacent etalons might seem daunting, this can be achieved by manufacturing a single large format etalon to uniform thickness (a $30\,nm$ is $\sim\frac{\lambda}{15}$, and can be easily verified using a laser interferometer), and cut it into several smaller etalons. Thickness variation between etalons can then be established at the nanometer scale by depositing a controlled thickness of SiO${\rm _2}$ on each etalon (private correspondence with LightMachinery).

\subsection{Separating Interference orders}
The beam transmitted by each etalon is a polychromatic beam with each wavelength component (\textit{i.e.} each interference order, see Eq.\,5) separated from the next order by the etalon's FSR, see Figs. \ref{fig:singleFPIResponse}-\ref{fig:FPIArrayResponse}. In order to obtain full spectral information (\textit{i.e.,} lift the interference order degeneracy), we need to separate the overlapping interference orders of each etalon in the array. To achieve that, we use an external long-slit spectrograph with a resolving power equal to or higher than the FSR of the etalons in the system. A FPI-based system effectively reduces the resolution requirement from a spectrograph by a factor equal to $\frac{\rm R_{inst}}{\rm FSR}\sim \rm N_{etalons}$, when compared to a stand-alone echelle spectrograph.

\begin{figure}[]
\centering
{\includegraphics[width=1\columnwidth]{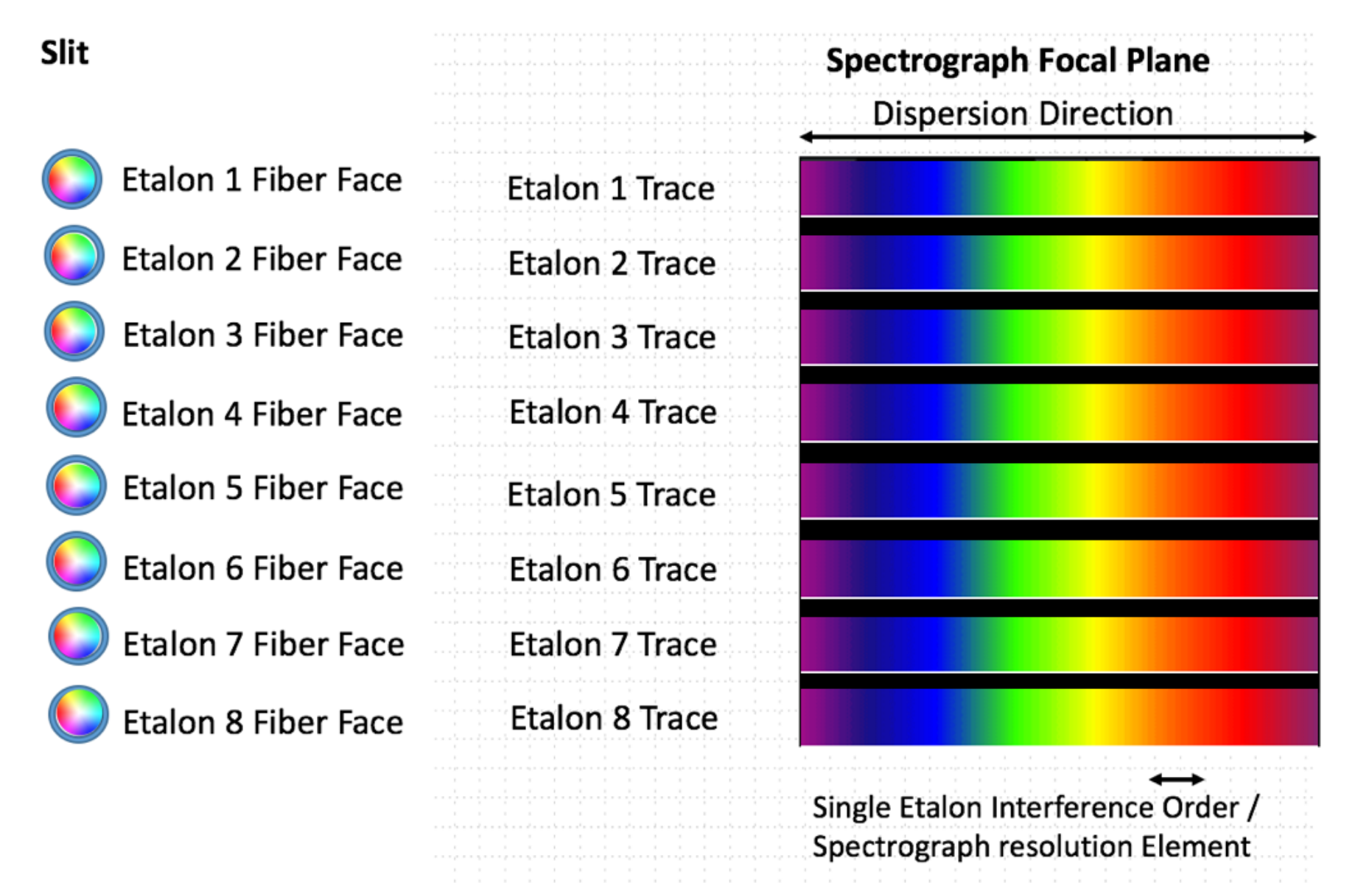}} 
\caption{Long-slit spectrograph Focal Plane. The transmitted beam from each etalon in the array is imaged onto a single fiber. The Fibers are arranged in a long slit at the spectrograph input plane (Left). Each etalon transmitted beam is dispersed and imaged to a spectral trace (Right).}
\label{fig:SpecFocalPlane}
\end{figure}

For the FPI array whose transmission curve is shown in Fig. \ref{fig:FPIArrayResponse}, a spectrograph with spectral resolution $R\geq3.75\cdot10^4$ will allow us to separate the interference orders of each etalon in the array. Such a spectrograph is required to have a slit long enough to accommodate inputs from all etalons in the array simultaneously. We envision the transmitted beam from each etalon is imaged onto an optical fiber. The fibers are arranged in a long slit at the entrance of the spectrograph. For each etalon in the array, a spectra will be imaged on the spectrograph focal plane along the spectrograph dispersion direction, with the etalons separated along the spatial direction, see Fig. \ref{fig:SpecFocalPlane}.
Wavelength calibration of a FPI array instrument is achieved by illuminating the system with a narrow line source of known wavelengths. The position of a line image of known wavelength on the spectrograph focal plane for a specific FPI trace allows us to calibrate the FPIs in the chain despite the lower resolution of the long-slit spectrograph. Possible sources that can be used for wavelength calibration are Fabry Perot calibrators and laser frequency combs \citep[\textit{e.g.,}][]{Wildi2012,Murphy2012}.

The resolving power of our method is derived by combining several FPIs with theoretical spectral resolving powers of $R\sim\mathrm{O}(10^6)$, with an external spectrograph with a resolving power significantly lower to separate interference orders. In analogy with classical echelle spectrographs in which a cross disperser (usually a prism or a grating working at $m=1$) is used to separate overlapping orders, in our case, the external spectrograph is the FPI array cross disperser.

While one can build a custom cross dispersing spectrograph tailored to the FPI array parameters, many existing spectrographs can perform order-sorting on the output of FPIs with minor modifications. One example is Hectochelle \citep{Szentgyorgyi2011}, an echelle spectrograph mounted on the MMT ${\rm 6.5\,m}$ telescope\footnote{The MMT aperture is most likely too small to allow ${\rm O_2}$ detection in terrestrial exoplanets' atmospheres, see \cite{Rodler2014}.}. Hectochelle can be fed by 240 fibers simultaneously (core diameter ${\rm \phi_{fiber}=250\mu m}$), and can achieve a resolution of ${\rm R\sim3.7\cdot10^4}$ over single, filter selected orders. Another example is G-CLEF \citep[][]{Szentgyorgyi2016,Ben-Ami2016}, a multi-mode echelle spectrograph currently being built for the ${\rm 25.4\,m}$ GMT. G-CLEF, when used in conjunction with  the MANy Instrument FibEr SysTem for the GMT \citep[MANIFEST;][]{Lawrence2016}, can accept over 40 fibers simultaneously, each with a core diameter of ${\rm \phi_{fiber}=300\mu m}$. In this operational mode, G-CLEF will deliver a spectral resolution of ${\rm R\sim3.75\cdot10^4}$.

ֿ\section{Implementation}
\begin{figure}[]
\centering
{\includegraphics[width=0.48\textwidth]{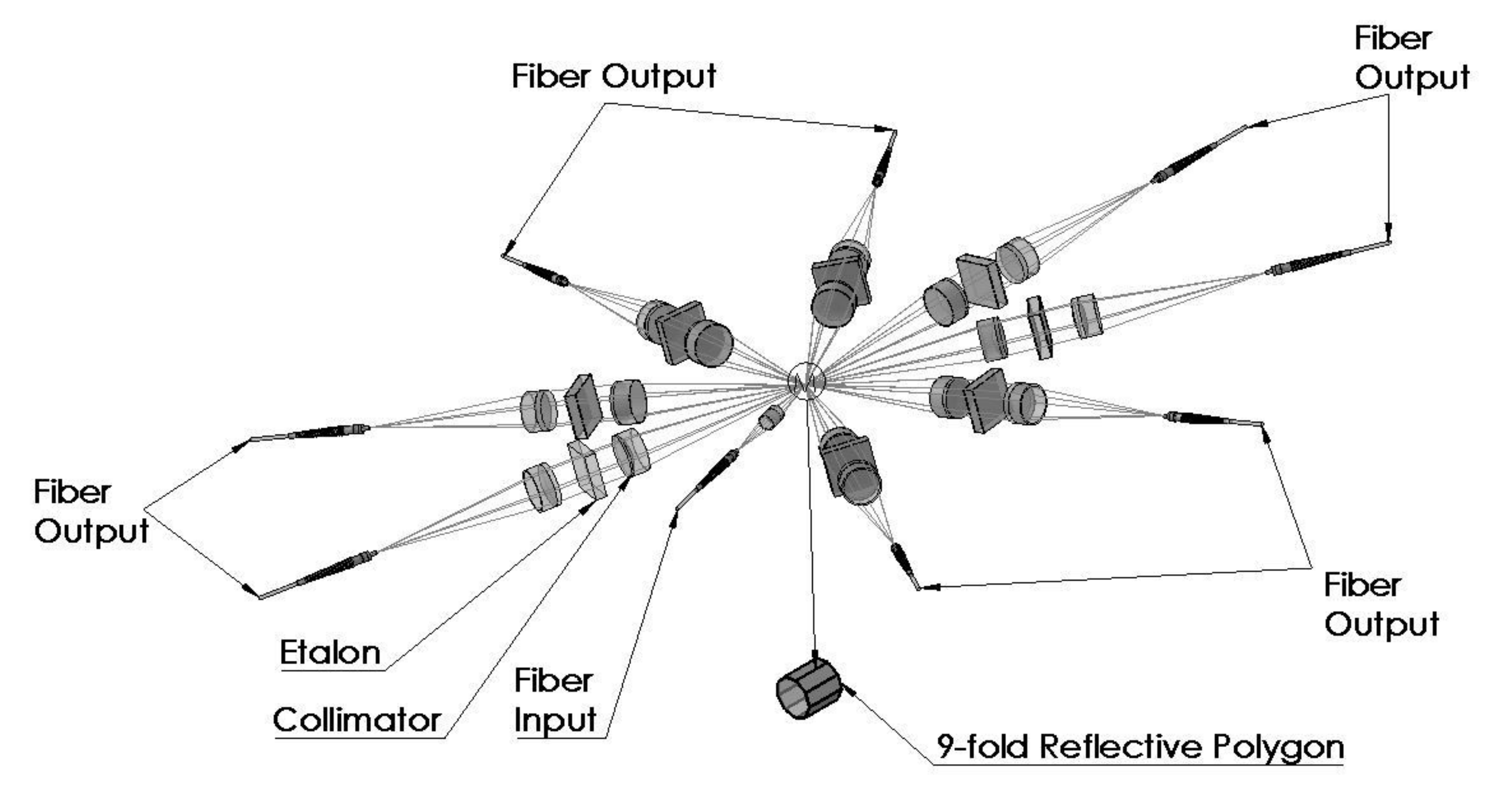}} 
\caption{An 8-fold Fabty Perot Interferometer Array.}
\label{fig:FIOS}
\end{figure}

In this section we discuss practical considerations that affect the design and performance of a FPI array similar to the one described in Section 2.2. In deriving numerical results, we assume the FPIs in the array are etalons, with parameters as given in Table 1. A computer-aided design (CAD) model of our  FPI array is shown in Fig. \ref{fig:FIOS}. 

The system is fed by a multi-mode optical fiber channeling light from the telescope image plane. The fiber face is imaged onto the reflective face of a 9-faceted reflective polygon. To allow propagation of the beam through the system, the incident angle on the polygon is $20^{\circ}$. The beam reflected by the polygon is collimated onto the first etalon in the chain. The angle of incidence (AoI) on the etalon is $\sim0.1^{\circ}$. The beam transmitted by the first etalon is focused onto an optical fiber while the beam reflected from the etalon is refocused onto the adjacent facet of the polygon mirror (\textit{i.e.,} for the reflected beam, the collimator works in double pass). The mirrored facet reflects the beam onto the next etalon in the array, with an identical setup (other than the etalon thickness, as described in Section 2.2)  - and so on $8$ times overall. The fibers terminate at the long slit of the cross dispersing spectrograph (see Section 2.3).  

To achieve maximum stability and minimize thermal and immersing index of refraction effects, the FPI array should be mounted inside a  thermally stabilized vacuum cell. Given a fused silica etalon with a thickness of $9.77\,$mm, we derive a thermal tuning sensitivity of ${\rm 4.9}$pm/K (\textit{i.e.,} the transmission peak shifts by $4.9\,$pm per $\Delta T=1^{\circ}$K). Limiting a thermally induced shift of the transmission comb in wavelength space to less than ${\rm 10\%}$ of the etalon theoretical resolution of ${\rm \sim9.2\cdot10^5}$ requires thermal stability of ${\rm \Delta T=\pm20\,mK}$ or better. Such thermal stability has been demonstrated on several of the leading HiRes spectrographs \citep[e.g.,][]{Becerril2010,Mueller2016}. The FPI array described here, including auxiliary optics, can fit in a $1\mathrm{m}\times1\mathrm{m}\times10\mathrm{cm}$ vacuum chamber.

\subsection{FPI Plate Size}
An advantage of using FPIs to achieve high resolving power is their small size compared to echelle gratings. We estimate the minimal FPI plate/facet dimensions required to conserve telescope \'etendue.
The \'etendue of a FPI depends on its plate/facet area ${\rm S}$ and spectral resolution ${\rm R}$ such that 
\begin{equation}
 {\rm U_{FPI}}=\frac{\rm k2\pi S}{\rm R} 
\end{equation}
where ${\rm k}$ is a numerical parameter of order unity \citep{Vaughan1989}. 
Equating the FPI \'etendue to that of a telescope with an aperture diameter ${\rm d_{tel}}$, we derive the relation
\begin{equation}
\frac{\rm \psi d_{tel}}{\rm 2\sqrt{2k/R}}={\rm d_{col}}    
\end{equation}
\noindent
where ${\rm d_{col}}$ is the diameter of the collimator feeding the FPI - and therefore sets ${\rm d_{FPI}}$, the FPI minimum clear aperture  (CA; The unobscured portion of an optic).

We can rearrange the above equation to derive a relation similar to Eq. 1 for a FPI:
\begin{equation}
{\rm R}={\rm 8k\cdot\left(\frac{d_{col}}{\psi d_{tel}}\right)^2}    
\end{equation}
\noindent which clearly illustrates the benefits of a FPI in achieving extreme HiRes compared to an echelle spectrograph.

Using Eq.\,11, we find ${\rm d_{FPI} \sim7.5\,mm}$ is the minimum  plate size for achieving ${\rm R=9.2\cdot10^5}$ using a FPI on a ${\rm D=6.5\,m}$ telescope with seeing conditions of ${\rm \psi=0.7''}$. In the case of a ${\rm D=25.4\,m}$ telescope with similar seeing conditions (\textit{e.g.,} the GMT), the minimum required plate diameter is ${\rm d_{FPI}\sim30\,mm}$. An echelle grating instrument operating at $\tan\delta=4$ with similar resolving power and seeing conditions would require a beam diameter of $\sim2.5\,$m  for a ${\rm D=6.5\,m}$ telescope, and a beam diameter of ${\rm \sim9.7\,}$m for a ${\rm D=25.4\,m}$ telescope.
The above analysis assumes normal incident on the FPI plates. For our application, the etalon facet is required to allow propagation of the beam as it is reflected back and forth between the etalon facets. For the small AoI considered in this work, the etalon facet size is increased by less than $15\%$, assuming $\sim50$ reflections are required to guarantee intensity attenuation of less than $0.1\%$.

\subsection{Throughput, Spectral Purity and Finesse}
In the example system presented, the theoretical spectral resolving power of each etalon is $\times3$ higher than the system sampling frequency. A priori, this seems like an unnecessary requirement which will result in undersampling, as well as a decrease in the system throughput. The reasoning behind setting the inherent resolving power at such a high value originates in the interplay between throughput and spectral purity. When determining the Surface Reflectivity Intensity, one sets the throughput and spectral purity of the system. A low SRI will broaden the transmission function around peak wavelengths, increasing the system throughput. However, this will cause a reduction in the specrtal purity - as the FPI contrast decreases for lower values of SRI (\textit{i.e.,} the amount of photons transmitted by each FPI with wavelengths outside of the FWHM of a given peak will increase).

\begin{figure}[]
\centering
{\includegraphics[width=0.99\columnwidth]{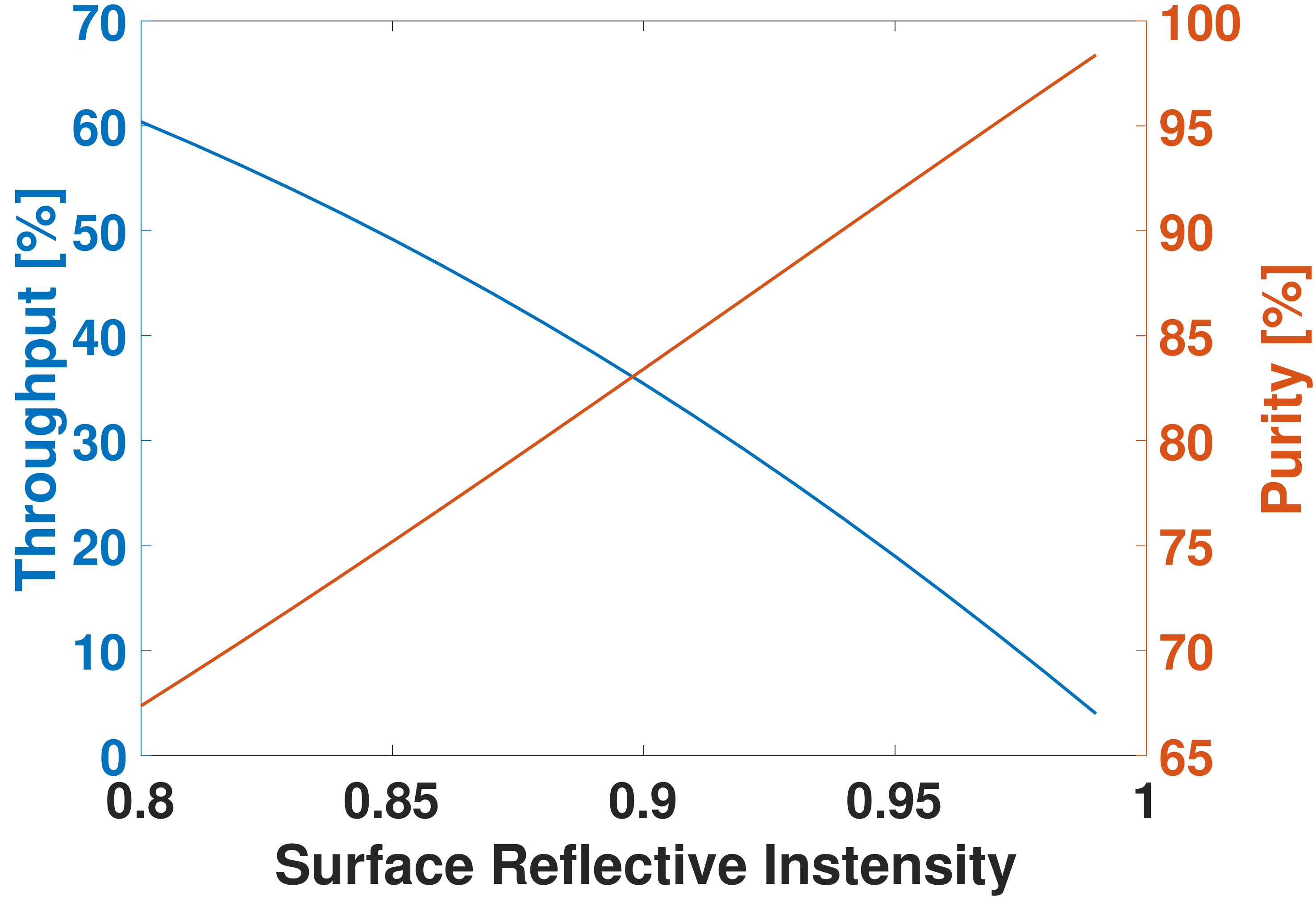}} 
\caption{Throughput/Spectral Purity analysis results. We vary the Surface Reflective Intensity between $0.80$ and $0.99$, and calculate the throughput (purity) within $\pm\frac{\lambda}{2R}$ ($\pm\frac{\lambda}{2\cdot FSR}$) of peak transmission.}
\label{fig:purityAnalysis}
\end{figure}

We illustrate the above by performing a throughput/purity analysis. We calculate the throughput of the FPI array across single resolution elements. The spectral purity of the recorded spectrum is determined by integrating over the out-of-peak transmission within a resolution element of the cross dispersing spectrograph (which is of the order of the FPI FSR, see Section 2.3). We assume FPI parameters as given in Table 1, while varying the SRI from $0.8$ to $0.99$. Our results are shown in Fig. \ref{fig:purityAnalysis}. 

As can be seen, a SRI of $R=0.88$ will result in signal purity of $\sim80\%$ similar to the one achieved in grating based spectrographs following the Rayleigh criterion \citep[\textit{e.g.,}][]{Schroeder2000}. The throughput of the FPI array is $\sim40\%$, averaged over all FPIs in the system, assuming $1\%$ loss due to reflection losses from auxiliary optics per additional FPI in the chain. While one can achieve higher spectral purity by increasing the SRI, this will cause an unnecessary decrease in the system throughput. A method to increase both throughput and purity is using dualons instead of etalons, see discussion in Section 5.

\subsection{Plate imperfections}
\begin{figure}[]
\centering
{\includegraphics[width=1\columnwidth]{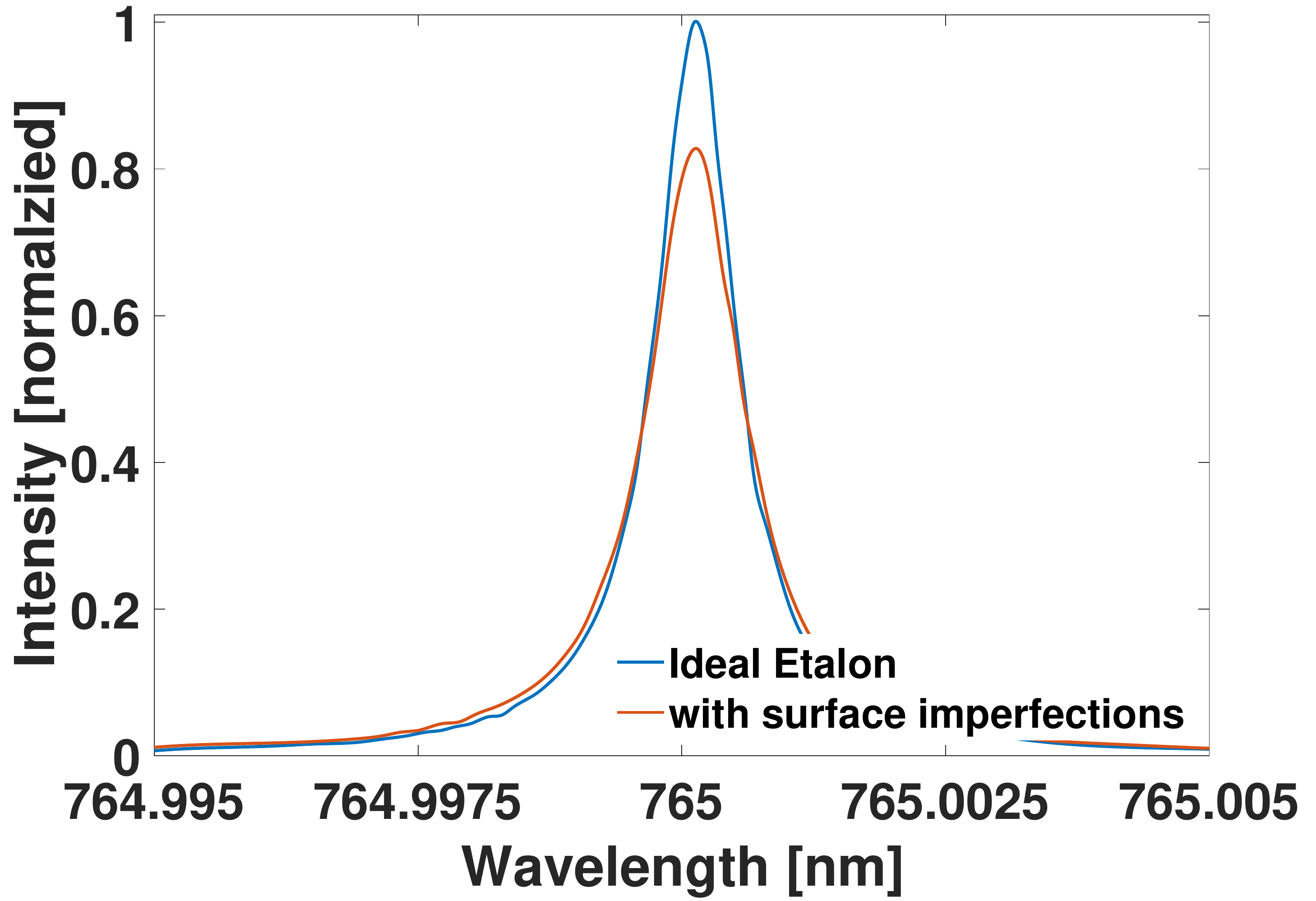}} 
\caption{The effect of surface imperfections on a FPI performance. Local imperfections are equivalent to local thickness variations that cause a broadening of the transmission profile, as well as a reduction in peak transmission.}
\label{fig:surfaceImperfections}
\end{figure}

The analysis presented in Section 2 assumes an ideal FPI. The two reflective facets of each etalon were assumed to be perfectly smooth, flat, and parallel. In reality, manufacturing imperfections and alignment errors should be taken into account. A local imperfection in a FPI reflective surface flatness, or an inherent tilt between the two reflective surfaces, are equivalent to a local change in the FPI thickness. The phase term $\phi$ (Eq. 4) will have a spatial dependency and will cause a broadening of the transmission peaks and a spectral resolving power degradation. The resultant instrument transmission profile is the convolution of the ideal Airy disk, $A(\phi)$, with the surface distribution function $D(\delta\phi)$ \citep[][and reference therein]{Vaughan1989}. We focus on three imperfections: 
\begin{itemize}
\item ${\it Residual~power~between~reflective~facets}$. This is typically quantified by the amount of spherical curvature between facets, and will result in a constant distribution of $\delta\phi$ between two extreme phase lags. The phase variation $\delta\phi$ will depend on the etalon diameter through the sag equation. 
\item ${\it Local~random~imperfections}$, which we quantify using a normal distribution with a RMS equal to the surface roughness.
\item ${\it A~tilt~between~the~two~reflective~surfaces}$, which will result in a linear increase in thickness as a function of plate diameter. From the phase equation, one can derive an hemispherical distribution $D(\delta\phi)$ for this case.
\end{itemize}
The expected performance, given the parameters in Table 1, and assuming high precision tolerance parameters achieved today by leading etalon manufacturers (\textit{i.e.,} residual power of $\lambda/100$, mirror tilt error of  $0.05^{''}$, and surface irregularity of $1\,$nm rms) are shown in Fig. \ref{fig:surfaceImperfections}. We find that overall the spectral resolving power is reduced to ${\rm R\sim7\cdot10^5}$. In addition, the peak transmission falls to $\sim83\%$. 

We can cast the discussion above in terms of an effective finesse. Given the imperfections scales listed above, the system finesse drops from a theoretical value of $\mathcal{F}=24.56$  to an effective value of $\mathcal{F}=19.1$ \citep[\textit{e.g.,}][]{Vaughan1989}.

\subsection{Beam Collimation and Finite Beam Divergence}
An additional degradation of the FPI array spectral resolving power and transmission efficiency is introduced by finite divergence of the collimated beam. This is not due to imperfections in the FPIs, but a property of the setup.

For a monochromatic beam, it is straightforward to characterize collimation effects in terms of the pinhole finesse \citep[\textit{e.g.,}][]{Guenther1990}:
\begin{equation}
    \mathcal{F}_{ph}={\rm \frac{4\lambda f_{col}^2}{D_p^2d}}
\end{equation}
\noindent with $\rm f_{col}$ the collimator focal length, and ${\rm D_p}$ the object size (\textit{i.e.,} the fiber core diameter in our case). For the system shown in Fig. \ref{fig:FIOS} (see parameters below), we derive  pinhole finesse value of $\mathcal{F}_{ph}\sim200$.

The pinhole finesse does not consider chromatic effects, or non normal incident angles. As imperfect collimation is often the limiting factor in system performance, we perform a rigorous analysis of its effects for the system shown in Fig. \ref{fig:FIOS}. For an optimized fiber fed system, a finite divergence in the incident angle of the collimated beam ${\rm \theta}$, is expected at a level of ${\rm \delta\theta=D_{p}/2f_{col}}$, where ${\rm D_{p}}$ is the fiber diameter. Rays passing through a FPI at angles other than ${\rm \theta}$ will have a phase lag of ${\rm \delta\phi=\frac{4\pi}{\lambda}dn_{\lambda}\sin\theta\delta\theta}$ with respect to the chief ray. This will result in additional broadening of the transmission peaks. While the real distribution of ray angles depends on the fiber far field, as well as the exact collimator surface figure, for the present study we assume a uniform distribution with a maximum deviation of ${\rm \delta\theta_{max}=\pm\Phi_{fiber}/2f_{col}}$. We further assume a uniform distribution function ${\rm D(\delta\phi)}$ within
\begin{equation}
    {\rm \delta\phi_{max}=\frac{4\pi}{\lambda}dn_{\lambda}\cdot(\sin(\theta\pm\frac{\Phi_{fiber}}{2f_{col}})-\sin\theta)}
\end{equation}
The phase lag distribution can then be convolved with the Airy function to assess the effective transmission profile of the FPIs. 

\begin{figure}[]
\centering
{\includegraphics[width=0.885\columnwidth]{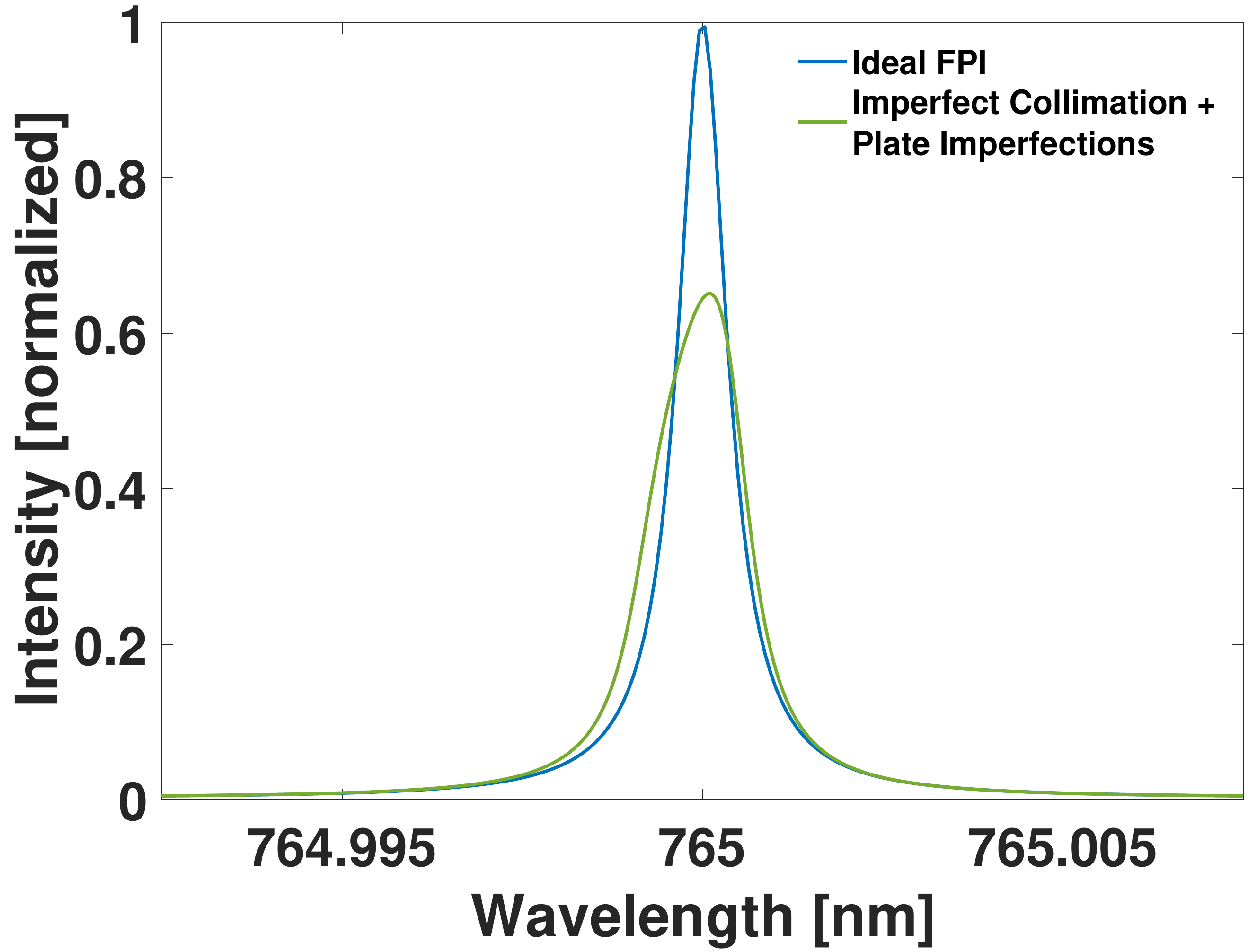}} 
\caption{Top: Deviations from ideal collimation will cause a degradation of resolving power, throughput, and contrast. In addition, the transmission profile is now skewed towards higher values of the feed angle. The red curve takes into account plates imperfections as discussed in Section $3.3$. The expected angle distribution is derived from an optical model of the system presented in Fig. \ref{fig:FIOS}.}
\label{fig:colimationEffect}
\end{figure}

To get a quantitative estimate of the degradation in PSF/collimation as the beam propagates through the optical train, we model the system shown in Fig. \ref{fig:FIOS} using the commercial optical design software Zemax. We estimate the expected angle distribution for rays reflected between facets at the last etalon in the chain, assuming the collimators are aspheric doublets with ${\rm f_{col}=250\,}$mm and an input fiber core diameter of ${\rm D_{p}=300\,\mu m}$. We derive the beam divergence assuming a polychromatic beam with $\lambda=750--780\,nm$. The results are shown in Fig. \ref{fig:colimationEffect}. We find an additional reduction in throughput, as well as a small contrast drop to $\frac{I_{max}}{I_{min}}\sim175$ (\textit{i.e.,} a degradation of $\sim0.6db$). In addition, we find the inherent FPI resolution drops to ${\rm R\sim5\cdot10^5}$, and the transmission profile becomes skewed. This increases for higher values of ${\rm \theta}$, readily apparent when examining the phase lag equation.

\begin{figure}[]
\centering
{\includegraphics[width=0.975\columnwidth]{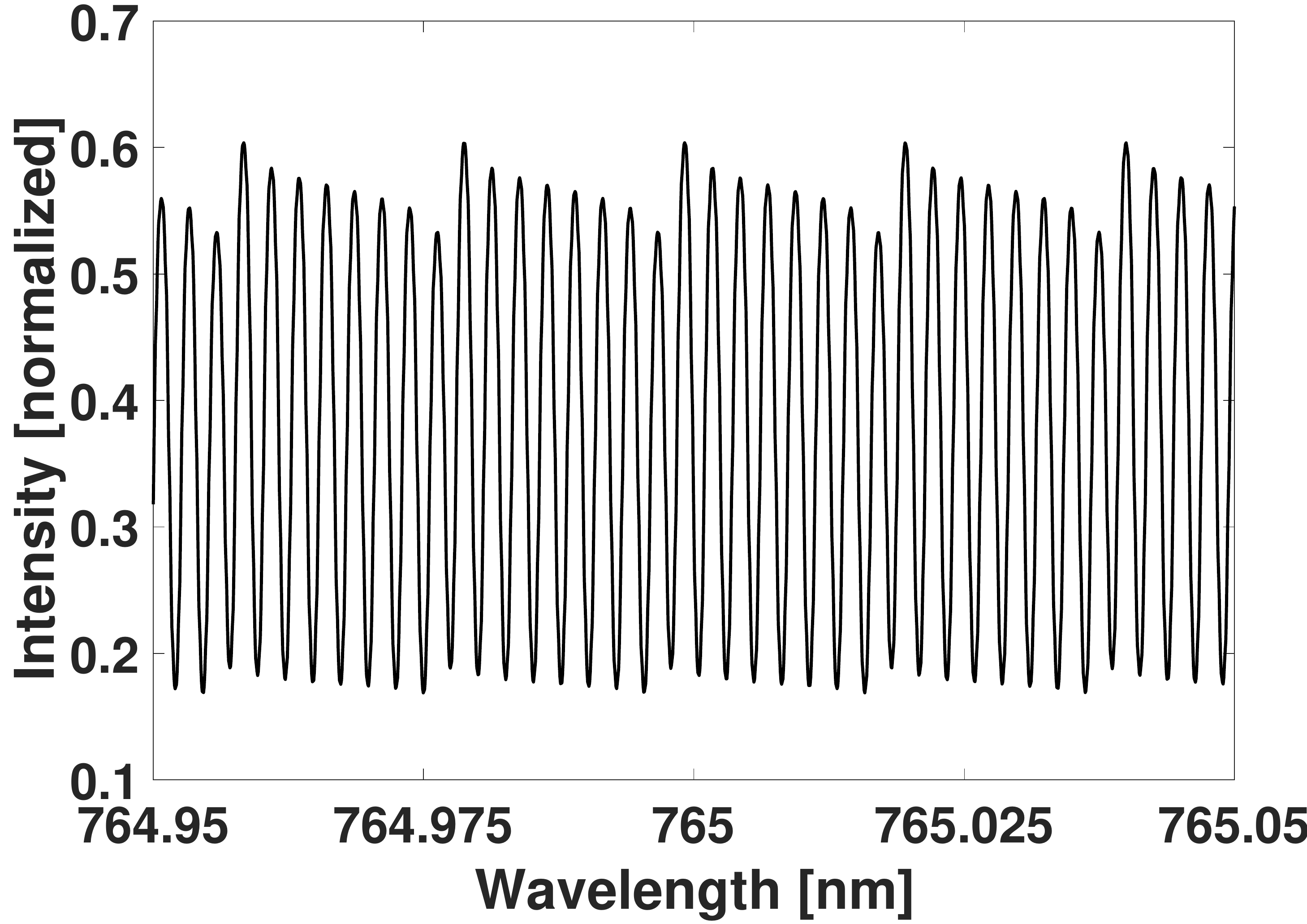}}
\caption{Efficiency across several interference orders, with plate imperfections and deviation from ideal collimation taken into account.}
\label{fig:overallEfficiency}
\end{figure}

Figure \ref{fig:overallEfficiency} shows estimated efficiencies of the full array across several interference orders, when including plate imperfections and finite beam divergence. The example FPI array has an efficiency of $\sim40\%$, resolving power of ${\rm \sim5\cdot10^5}$, and sampling frequency of $\sim3\cdot10^5$. Various methods to reduce resolution degradation and reduced throughput due to surface imperfections and imperfect collimation are discussed in Section 5. 

\section{Simulation}
In this section we address the question of whether a FPI array can improve the efficiency of ${\rm O_2}$ detection in exoplanets atmospheres. In particular, we compare the performance of the example FPI array with ${\rm R=3\cdot10^5}$ and ${\rm R=5\cdot10^5}$ to performance estimates of echelle spectrographs on ELTs\footnote{The etalons in the ${\rm R=5\cdot10^5}$ array are similar to the example etalons, but with a distance step of $20$\,nm between adjacent etalons in the chain. In addition, there are 13 etalons in the chain.}. \cite{Rodler2014} find that in the optimal case of an Earth twin orbiting in the habitable zone of an M4V star ${\rm 5\,pc}$ away, it would require 34 transits to obtain a ${\rm 3\sigma}$ detection of ${\rm O_2}$ with an instrument like G-CLEF on the GMT. 

We follow the same methodology presented in \cite{Rodler2014} and L${\rm  \acute{o}}$pez-Morales et al. (2018; In Prep), to generate model spectra following the equations
\begin{equation}
  {\rm a} = {\rm (1+v_{*}c^{-1})S+(1+(v_{*}+v_{pl})c^{-1}\epsilon^{-1})P},
\end{equation}
\begin{equation}
    {\rm \tilde{C}} = {\rm (a(1+\epsilon^{-1})^{-1}T)} ,
\end{equation} 
\begin{equation}
{\rm }C = {\rm \tilde{C}} \times G       
\end{equation}
\noindent
where ${\rm S}$ is the spectrum of the host star, ${\rm P}$ is the transmission spectrum of the planet, ${\rm T}$ is Earth's telluric spectrum, ${\rm \tilde{C}}$ is the combined star, planet, and telluric spectrum, and ${\rm G}$ is the instrumental profile. The parameters, ${\rm v_{*}}$ and ${\rm v_{pl}}$ are, respectively, the velocity of the planet host star with respect to Earth, and the instantaneous radial velocity of the planet with respect to its host star (which is ${\rm \sim0\,km\,s^{-1}}$ at primary transit). The variable $\epsilon$ is the ratio of the area of the stellar disk and the area of the atmospheric ring of the planet.

For the stellar spectrum ${\rm S}$ we use a M4V star model with effective temperature ${\rm T_{eff} = 3000~K}$, gravity ${\rm logg = 4.5~dex}$, and solar abundances from the \cite{Husser2013} library of high-resolution spectra calculated with the PHOENIX \citep{Brott2005}. For Earth's telluric spectrum ${\rm T}$, we use the spectrum described in Section 2.1.2 of \cite{Rodler2014}. The exoplanet transmission spectrum ${\rm P}$ was generated using the Reference Forward Model \citep[RFM;][]{Dudhia2017}. In our simulations the RFM is driven by the HITRAN 2012 spectral database \citep{Rothman2013}. We use the US standard atmosphere (1976)\footnote{https://ccmc.gsfc.nasa.gov/modelweb/atmos/us$\_$standard.html} to simulate temperature, pressure, and molecular oxygen profiles from the surface of the planet up to $100\,$km at $2\,$km intervals (layers). We generate the transmission spectrum of the exoplanet by integrating the transmission of each layer. To account for refraction in the exoplanet atmosphere, we do not include the first three layers of the model, between $0--6$\,km, in the integration \citep{Rodler2014}. We assume {$\rm v_{*}=25\,$km\,s$^{-1}$} and ${\rm \epsilon=35,000}$ following \citep{Rodler2014}.

We estimate the number of transits needed to achieve a ${\rm 3\sigma}$ detection of ${\rm O_2}$ with our model FPI array. We also estimate the number of transits needed for such detection with G-CLEF on the GMT, to establish and confirm a baseline with \cite{Rodler2014}. For the G-CLEF model we use a spectral resolution of ${\rm R=10^5}$, and assume a Gaussian instrument profile with a FWHM equal to the instrument spectral resolution. We then cast the convolved spectrum onto a pixel grid with a velocity resolution of ${\rm 0.75\,km\,s^{-1}}$, which is the planned grid scale for G-CLEF \citep{Szentgyorgyi2012}.

For the FPI array models we assume resolutions of ${\rm R=3\cdot10^5}$ and ${\rm R=5\cdot10^5}$. To simulate the spectrum observed through the array, we take several additional steps: 
\begin{itemize}
\item The spectrum $\tilde{C}$ is passed through a filter function based on the transmission properties of each etalon in the array (see Figs. \ref{fig:colimationEffect} and \ref{fig:overallEfficiency}).
\item The transmitted output of each etalon is convolved with a Gaussian with a FWHM equivalent to the cross dispersing spectrograph resolving power of $R=3.75\cdot10^4$, see Section 2.3.
\item The transmitted spectrum from each etalon is cast onto a pixel grid with a velocity resolution of $0.75\,$km\,s$^{-1}$.
\item For each spectrum, a wavelength solution is applied according to peak transmissions of the given etalon in the array. At this point, we produced $8\,(13)$ spectra with a spectral resolution of ${\rm \frac{\lambda}{\delta\lambda}\sim3\,(5)\,\cdot10^5}$. Each spectrum has a sampling frequency of ${\rm \frac{\lambda}{\Delta\lambda}\sim3.75\cdot10^4}$, with the transmission comb shifted by ${\rm \frac{\lambda}{\delta\lambda}\sim3\,(5)\,\cdot10^5}$ between etalons in the array.
\item The 8 spectra are combined to a single spectrum with a spectral resolution of ${\rm R\sim3\,(5)\,\cdot10^5}$.
\end{itemize}

After obtaining the model spectra $C$ for each case (\textit{i.e.,} spectra directly imaged by G-CLEF or spectra transmitted through the FPI array and cross-dispersed by an external spectrograph), we estimate the photon flux per transit. We assume a ${\rm 2.1\,}$hours transit duration, and a duty cycle of $0.98$ (in the M4V case, smearing of the absorption lines due to radial velocity shifts of the exoplanet limits the duty cycle - see \citealt{Rodler2014} and discussion in L${\rm \acute{o}}$pez-Morales et al. 2018; In Prep). For all cases, we assume the instrument is mounted on the GMT (\textit{i.e.,} a collecting area of $368$m$^2$). Finally, we set the instrument efficiencies to be identical in the three cases and equal to G-CLEF's end-to-end efficiency, see discussion in Section 5.

At this point we add noise terms. Our noise model assumes white noise following a Poisson distribution for each pixel photon count and red noise equivalent to ${\rm 20\%}$ of the white noise for the ${\rm R=10^5}$ case \citep{Rodler2014}. For each spectral resolution investigated we generate 1000 model spectra with randomized noise values.

The final step in the simulation is a bootstrap analysis in which we combine $N$ randomly selected model transit spectra for a specific resolution, with $N$ being the number of observed transits. We reduce each sample to obtain a residual spectrum such that: 
\begin{equation}
P_{residual}=\frac{C_{inTransit}}{C^{noiseFree}_{outOfTransit}}   
\end{equation}
\noindent where $C^{noiseFree}_{outOfTransit}$ is the observed out-of-transit spectrum in the infinite signal-to-noise ratio case (in the noise-free case of $C_{inTransit}$, the residual spectrum would have contained only the exoplanet transmission spectrum). We cross correlate $P_{residual}$ with a template exoplanet transmission spectrum. 
We examine the cross correlation function and determines its peak significance in $\sigma$, assuming its peak is at $25\,\mathrm{km\,s^{-1}}$, the standard deviation of the cross correlation function over all possible velocities (if the peak is at a different velocity, we assign zero significance). We repeat the process $500$ times for each spectral resolution - transit number combination.

\begin{figure}[]
\centering
{\includegraphics[width=1\columnwidth]{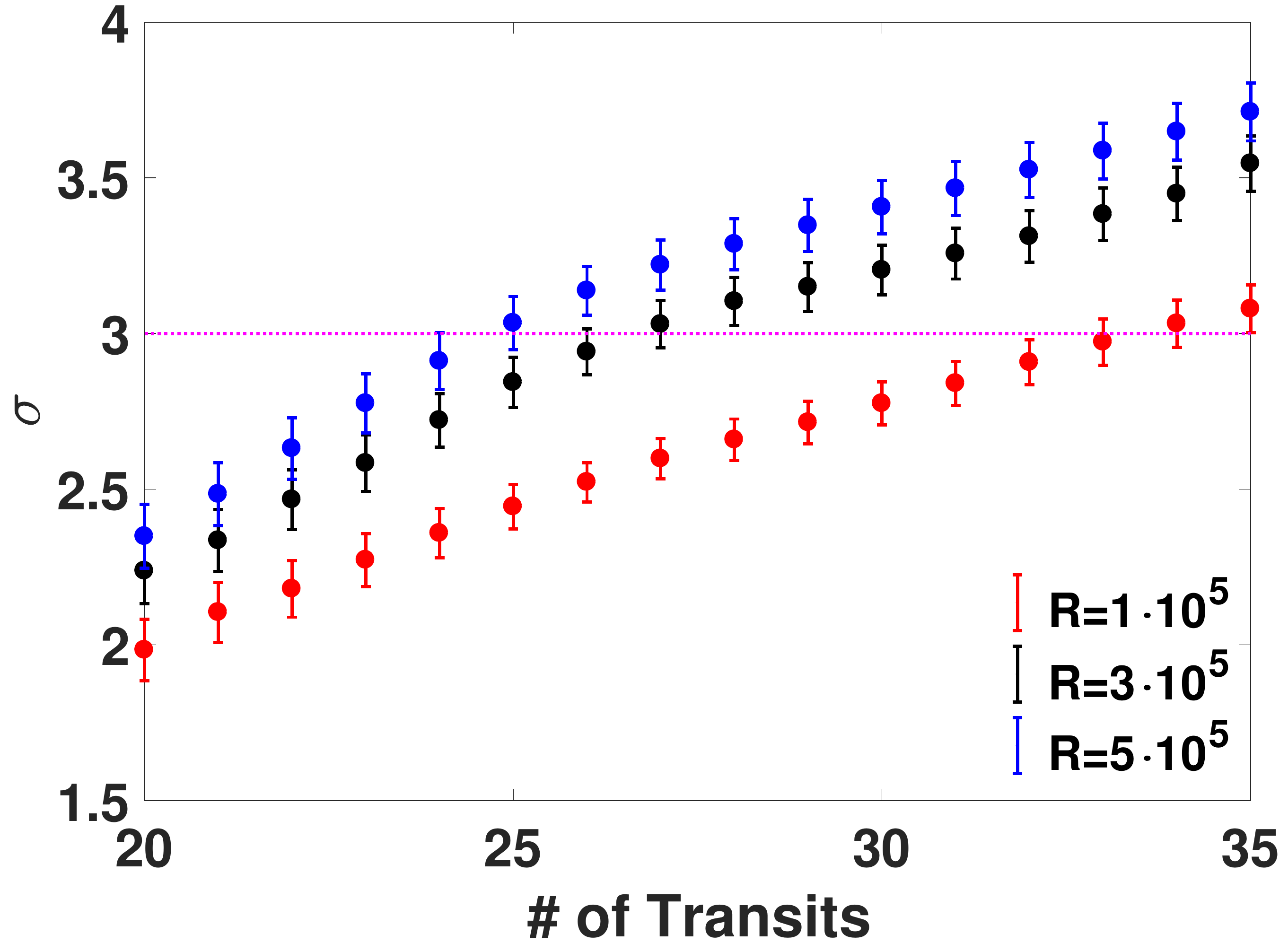}} 
\caption{Bootstrap analysis Results. The number of transits needed for a $3\sigma$ detection drops by $\sim7-10$ transits when observing with the exemplar FPI arrays.}
\label{fig:transitEstimates}
\end{figure}

Our analysis results are shown in Fig. \ref{fig:transitEstimates}, with the error bars indicating the standard deviation of the 500 instances. For the ${\rm R=10^5}$ case, we reproduce the results of \cite{Rodler2014}, showing that ${\rm \sim33-34}$ transits are needed to detect ${\rm O_2}$ at a $3\sigma$ level when observing with the GMT. The simulations also indicate that the number of transits needed for a $3\sigma$ detection drops to $\sim26-27$ when observing with a FPI array at ${\rm R=3\cdot10^5}$, and to $\sim24-25$ when observing with a FPI array at ${\rm R=5\cdot10^5}$. Overall, the number of transits needed for a $3\sigma$ detection drops by $25\%-35\%$ when observing with the example FPI array. This, in turn, can lead to a reduction of $\sim3-4$ years in an observing program targeting molecular oxygen detection, assuming one in every nine transits can be observed \citep{Rodler2014}. Our simulations illustrate how a FPI array instrument of modest dimensions can record extreme HiRes spectra in natural seeing conditions. In addition, assuming similar throughput, our simulations confirm the benefits of increasing the spectral resolution for molecular oxygen detection in transmission spectroscopy of terrestrial exoplanets.

\section{Summary and Discussion}
In this paper we presented a novel method for extreme HiRes spectroscopy based on the transmission and reflection properties of FPIs. Our method utilizes a FPI array to generate a transmission comb with a spectral resolution and sampling frequency well in excess of $\rm R\sim10^5$, the typical resolution of modern HiRes echelle spectrographs. We presented an initial design of such an array, choosing etalons as the main components of the array and relying on modern manufacturing techniques, which allow control of an etalon thickness at a few nanometers level. 

In a FPI array instrument, the beam from a source at the telescope focal plane is collimated onto the 1st etalon in the array. Due to the optical path difference between reflected beams at the etalon boundaries, rays of specific wavelengths will be transmitted while others will be reflected. The beam reflected from the  etalon is collimated onto the next etalon in the array, designed to have a slightly different thickness so that its transmission function is slightly shifted in wavelength space. This is repeated for ${\rm N}$ etalons in the array, with ${\rm N}$ dictated by the ratio of the instrument target resolution to the free spectral range of a single etalon. The beam transmitted by each etalon is imaged along the slit of an external spectrograph, thus retaining discrimination between etalons in the chain.
The external spectrograph, with a spectral resolution similar to the etalons free spectral range, is used to separate interference orders from each etalon, thus achieving full spectral information for each of the ${\rm N}$ transmitted beams.

We analyzed the transmission properties of a FPI array, from a single etalon to the response of an ${\rm N}$-fold array. We showed that the optics size of a FPI array is significantly smaller than the optics needed for an echelle spectrograph of similar resolution, making such an instrument feasible for the next generation ELTs when operating at seeing limited conditions below $\lambda\sim1\,\mu$m. We also suggested a criterion to determine the etalon surface reflection intensity, similar to the Rayleigh criterion for an long-slit spectrograph resolution. We determined environmental conditions needed to achieve a stable response from the array, and discussed the effects of manufacturing imperfections and imperfect collimation, showing that the desired spectral resolution can be achieved when taken into account.

Throughout the paper,we focused the discussion on ${\rm O_2}$ detection in the atmosphere of transiting exoplanets. We presented initial simulation results of FPI array observations of a transiting Earth twin in the habitable zone of an M4 star. Our results indicate that such a FPI array instrument with ${\rm R=3-5\cdot10^5}$ will allow us to reduce the number of transits needed for a ${\rm 3\sigma}$ detection by roughly ${\rm \sim25-35\%}$ when compared to a similar observing program using an echelle spectrograph with a resolution of ${\rm R\sim10^5}$. This, in turn, can lead to a reduction of $\sim3-4$ years in an observing program targeting molecular oxygen detection.

In the simulations we assumed similar efficiencies for all instruments. This can be justified if the required resolution for order separation allows us to avoid the use of a pupil slicer, and increase the size of the fiber core diameter feeding the spectrograph. In the case of G-CLEF, the efficiency difference between its highest and lowest resolution modes can be up to a factor of $\sim2$. 

\begin{figure}[h!]
\centering
{\includegraphics[width=0.99\columnwidth]{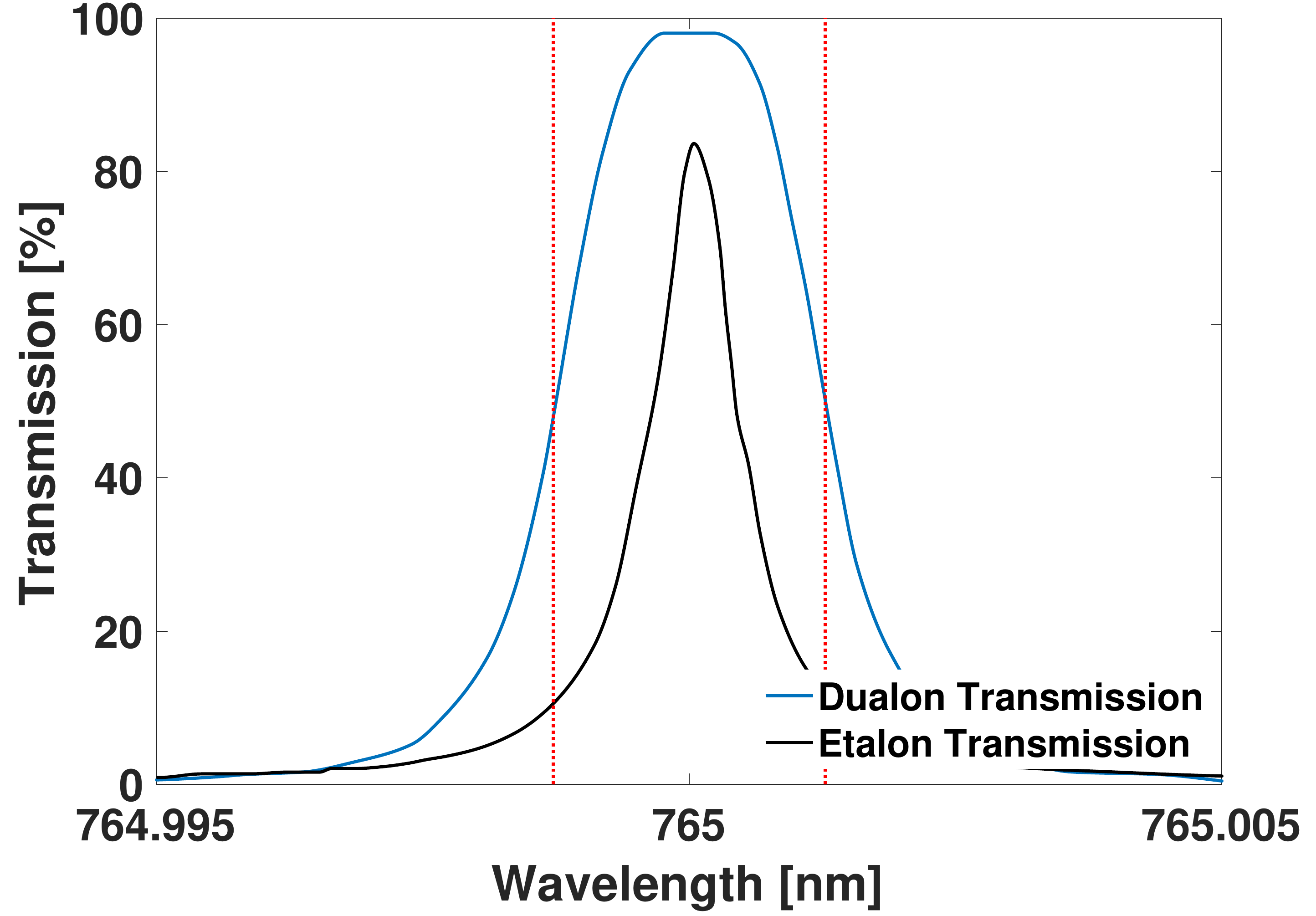}} 
\caption{Dualon Transmission. Pick transmission of a dualon has a flat top. The overall transmission of the dualon across a single resolution element is a factor of $2.2$ times greater than that of the etalon. Plates/Facets imperfections are taken into account in both cases.}
\label{fig:DualonComparisson}
\end{figure}

The final instrument setup needs to take throughput considerations into account, and balance between the number of FPIs in the chain and properties of the external spectrograph. One way of increasing the system efficiency is by building a dedicated cross-dispersing spectrograph with dispersing elements, camera design, anti-reflection (AR) and reflection coatings, and detector unit optimized to achieve maximum efficiency at the narrow ${\rm O_2}$ A-band. If we assume, for example, that a modern echelle spectrograph has on the order of 20 air-to-glass interfaces, an optimization of anti-reflection coatings to the ${\rm O_2}$ A-band can result in a ${\rm \sim20\%}$ efficiency increase with respect to broad band AR coatings typical of today state-of-the-art instruments. In addition, if we can reduce the resolution required from the external spectrograph to a value low enough to consider dispersing elements other than echelle gratings, more efficient elements can be utilized. A second method for increasing efficiency is using dualons instead of etalons. Dualons are etalons with a ${\rm 3rd}$ reflective surface buried between the two boundary reflective surfaces. The transmission comb peaks of dualons have a top hat shape, and their suppression of out-of-band transmission is orders of magnitude higher. In the case shown in Fig. \ref{fig:DualonComparisson}, the overall transmission efficiency across a single resolution element from the dualon is  a factor of $2.2$ times greater than that of an etalon. A full analysis of the dualon case will be presented in a follow-up paper.

Finally, FPI array based instruments offering resolutions of ${\rm R>10^5}$ can be used for a range of science cases, from detection of the cosmological redshift drift (\textit{i.e.,} the Sandage-Loeb effect) through the study of dark matter using velocity dispersion measurements in dSph galaxies and the study of extreme low metallicity stars in the milky way halo \citep{Loeb1998,Beckers2007,Lazkoz2017,Mashonkina2017}. We are currently in the process of assembling a FPI array prototype in our labs, with parameters similar to the array presented in the work. We expect to publish our results in late 2018.

\acknowledgments
The authors thank the Brinson Foundation for their funding support for this project. SBA thank Ian Miller of LightMachinery for the fruitful discussion on the as built performance of etalons and the physical properties of dualons.


\end{document}